\documentclass[prd,superscriptaddress,nofootinbib,showpacs,preprint]{revtex4}

\usepackage{amsmath}
\usepackage{graphicx}
\usepackage{color} 

\newcommand{\vev}[1]{\langle #1 \rangle}

\begin{document}
\title{Spontaneous Parity Violation in a Supersymmetric Left-Right
Symmetric Model}
\author{Sudhanwa Patra}
\email{sudhakar@prl.res.in}
\affiliation{Physical Research Laboratory, Ahmedabad 380009, India}
\author{Anjishnu Sarkar}
\email{anjishnu@iopb.res.in}
\affiliation{Institute Of Physics, Bhubaneswar 751005, India}
\author{Utpal Sarkar}
\email{utpal@prl.res.in}
\affiliation{Physical Research Laboratory, Ahmedabad 380009, India}
\affiliation{Physics Department and McDonnell Center for Space Sciences,
Washington University, St. Louis, MO 63130, USA}
\author{Urjit A. Yajnik}
\email{yajnik@phy.iitb.ac.in}
\affiliation{Indian Institute of Technology, Bombay, Mumbai 400076, India}
\affiliation{Indian Institute of Technology, Gandhinagar, Ahmedabad 382424,
India}
\date{}

\begin{abstract}

We propose a novel implementation of spontaneous parity breaking in
supersymmetric left-right symmetric model, avoiding some of the problems 
encountered in previous studies. This implementation includes a bitriplet 
and a singlet, in addition to the bidoublets which extend the Higgs sector 
of the Minimal Supersymmetric Standard Model (MSSM). The supersymmetric 
vacua of this theory are shown to lead generically to spontaneous violation 
of parity, while preserving $R$ parity. The model is shown to reproduce 
the see-saw relation for vacuum expectation values, 
$v_L v_R \approx m_{EW}^2$ relating the new mass scales $v_L$, $v_R$
to the electroweak scale $m_{EW}$, just as in the non-supersymmetric
version. The scale $v_R$ determines the mass scale of heavy majorana 
neutrinos, which gets related to the obeserved neutrino masses through 
type II see-saw relation.

\end{abstract}

\maketitle

\section{Introduction}
\label{sec:intro} 

The standard model (SM) has been successful in explaining the strong, the 
weak and the electromagnetic interactions at currently accessible energies. 
The only obvious indication of physics beyond the SM seems to be the
observation of the neutrino mass. Yet there are several motivations to
look for a comprehensive solution to a variety of puzzles of the SM. 
Left-Right symmetric model \cite{Pati:1974yy, Mohapatra:1974gc, 
Senjanovic:1975rk, Mohapatra:1980qe, Deshpande:1990ip}
has since long received considerable attention as a simple extension of 
the SM. While chirality is an elegant ingredient of nature which prevents
unduly large masses for fermions, most of nature is left-right symmetric, 
suggesting the reasonable hypothesis that parity is only spontaneously broken,
a principle built into the left-right symmetric models. 
Due to inclusion of right handed neutrino states as a principle, such models 
provide a natural explanation for the smallness of
neutrino masses \cite{Fukuda:2001nk, Ahmad:2002jz, Ahmad:2002ka, 
Bahcall:2004mz} via see-saw mechanism \cite{Minkowski:1977sc, 
Gell-Mann:1980vs, Yanagida:1979as, Mohapatra:1979ia}.
This class of models also provides a natural embedding of electroweak
hypercharge, giving a physical explanation for the required extra $U(1)$ as
being generated by the difference between the 
baryon number ($B$) and the lepton number ($L$). Thus, $B-L$, the only exact 
global symmetry of SM becomes a gauge symmetry, ensuring its exact conservation,
in turn leading to several interesting consequences. 

The other extension of the standard model is the grand unified
theory, which unifies all the gauge groups into a single simple
group at very high energy
with only one gauge coupling constant to explain all the three
low energy interactions. The quark-lepton unification then 
predicts proton decay, charge quantization, etc. However, the
high energy scale leads to the gauge hierarchy problem, which
dictates the inclusion of supersymmetry as a key ingredient. In
order to protect the electroweak scale from the unification scales,
the minimal supersymmetric standard model (MSSM) would be the most
logical extension of the standard model. Its prediction of particles 
at energies accessible to current colliders makes the model of
immediate interest. The SM predictions
now get enriched by the additional predictions of supersymmetry,
but to prevent the unwanted predictions like proton decay, one
needs to impose the $R$-parity symmetry, defined in terms
of the gauged $(B-L)$ quantum number \cite{Martin:1997ns,Mohapatra:1986su},
as
\begin{equation} 
R = (-1)^{3(B-L)+2S}. 
\label{eq:rparity}
\end{equation}
In the class of supersymmetric left-right models where the parity
breakdown is signalled by the vacuum expectation values of triplet 
Higgs scalars, the $R$-parity is naturally
conserved and its origin gets related to the gauged $B-L$ symmetry
\cite{Aulakh:1997ba}.

The bare minimal anomaly free supersymmetric extension of the left-right
symmetric model with triplet Higgs bosons leads to several nettlesome 
obstructions which may be considered to be a guidance towards a unique 
consistent theory. 
One of the most important problems is the spontaneous breaking of
left-right symmetry \cite{Kuchimanchi:1993jg,Kuchimanchi:1995vk}, viz.,
all vacuum expectation values breaking $SU(2)_L$ are exactly equal
in magnitude to those breaking $SU(2)_R$, making the vacuum parity symmetric. 
There have been suggestions to solve this problem by introducing
additional fields, or higher dimensional operators, or by going
through a different symmetry breaking chain or breaking the
left-right symmetry along with the supersymmetry breaking 
\cite{Kuchimanchi:1993jg,Kuchimanchi:1995vk,Babu:2008ep,Aulakh:1998nn, 
Sarkar:2007er, Aulakh:1997vc, Aulakh:1997fq}. In some cases,
when the problem is cured through the introduction 
of a parity-odd singlet, the soft susy breaking terms lead to breaking of
electromagnetic charge invariance. A recent improvement \cite{Babu:2008ep}
using a parity even singlet may however deviate significantly from MSSM, 
and remains to be explored fully for its  phenomenological consistency.
Further, in the minimal SUSYLR model with minimal Higgs fields, which has
been studied extensively \cite{Aulakh:1997ba, Kuchimanchi:1993jg, 
Kuchimanchi:1995vk}, it has been found that global minimum of the Higgs 
potential is either charge violating or $R$-parity violating. In this 
article we propose yet another solution to the problem, which resembles
the non-supersymmetric solution, relating the vacuum expectation
values ($vev$s) of the left-handed and right-handed triplet
Higgs scalars to the Higgs bi-doublet $vev$ through a seesaw
relation. The left-right symmetry breaking scale thus
becomes inversely proportional to the left-handed triplet Higgs
scalar that gives the type II seesaw masses to the neutrinos. 
The novel feature consists in the introduction of a bitriplet Higgs and  
another Higgs singlet under left-right group. The vacuum that preserves 
both electric charge and R-parity can naturally be the global minimum of the
full potential. The most attractive feature of the present model
is that generically it does not allow a left-right symmetric vacuum,
though the latter appears as a single point within the flat direction 
of the minima respecting supersymmetry. When the flat direction is lifted
all the energy scales required to explain phenomenology result naturally. 
This model can be embedded in the minimal supersymmetric SO(10) grand 
unified theory. 

Section (\ref{sec:MSLRMrecap}) recapitulates the minimal supersymmetric
left-right model for completeness of the paper. In section
(\ref{sec:bitrip-sing}) we discuss the proposed new model of supersymmetry
having an additional bi-triplet and a singlet.
Section (\ref{sec:phenom}) discusses the phenomenology of this proposed model.
Finally, section (\ref{sec:conclusion}) gives the conclusion.

\section{Minimal Supersymmetric Left-Right Model: A recap}
\label{sec:MSLRMrecap}

In this section we briefly describe the minimal supersymmetric left-right
model. In the left-right symmetric models, it is assumed that
the MSSM gauge group $SU(3)_c \otimes SU(2)_L \otimes
U(1)_Y$ is enhanced at some higher energy, when the left-handed
and right-handed fermions are treated on equal footing. 
The minimal supersymmetric left-right (SUSYLR) model has the gauge group 
$SU(3)_{c}$ $\otimes ~SU(2)_{L}$ $\otimes ~SU(2)_{R}$ $\otimes ~U(1)_{B-L}$
which could emerge from a supersymmetric $SO(10)$ grand unified theory.
The model has three generations of quarks and leptons, and
their transformations are given by,
\begin{eqnarray}
Q = (3,2,1,1/3), &\qquad& Q^{c} = (3^{*},2,1,-1/3), \nonumber \\
L = (1,2,1,-1), &\qquad& L^{c} = (1,1,2,1),
\end{eqnarray}
where, the numbers in the brackets denote the quantum numbers under 
$SU(3)_{c}$, $SU(2)_{L}$, $SU(2)_{R}$, $U(1)_{B-L}$. We have omitted the
generation index for simplicity of notation. 

The left-right symmetry could be broken by either doublet Higgs
scalars or triplet Higgs scalar. It has been argued that
for a minimal choice of parameters, it is convenient to break the
group with a triplet Higgs scalar. We shall consider here the 
minimal Higgs sector, which consists of
\begin{eqnarray}
\Delta = (1,3,1,2),  & \qquad & \bar{\Delta} = (1,3,1,-2),
\nonumber \\
\Delta^{c} = (1,1,3,-2),  & \qquad & \bar{\Delta}^{c} = (1,1,3,2),
\nonumber \\
\Phi_{i} = (1,2,2^{*},0),  & \qquad & (i=1,2).
\end{eqnarray}
As pointed out in \cite{Aulakh:1997ba} the bidoublets are doubled to achieve
a nonvanishing Cabibbo-Kobayashi-Maskawa (CKM) quark mixing and the number 
of triplets is doubled for the sake of anomaly cancellation.
Left-right symmetry is implemented in these theories as a discrete parity
transformation as
\begin{eqnarray}
Q \longleftrightarrow Q_{c}^{*}, &
L\longleftrightarrow L_{c}^{*}, &
\Phi \longleftrightarrow \Phi^{\dagger} \nonumber \\ [0.3cm]
\Delta \longleftrightarrow {\Delta^{c}}^{*},&
\bar{\Delta} \longleftrightarrow \bar{\Delta}{^{c}}^{*}.
\end{eqnarray}
The superpotential for this theory is given by
\begin{eqnarray}
W &=& \nonumber Y^{(i)_{q}} Q^{T} \tau_{2} \Phi_{i} \tau_{2} Q^{c} 
  + Y^{(i)_{l}} L^{T} \tau_{2} \Phi_{i} \tau_{2} L^{c}\\ 
\nonumber 
&& + ~i (f L^{T} \tau_{2} \Delta L +f^{*} L{^{c}}^{T} \tau_{2} 
  \Delta^{c} L^{c} ) \\
&& + ~\mu_{\Delta} \textrm{Tr}(\Delta \bar{\Delta}) + \mu^{*}_{\Delta} 
  \textrm{Tr}(\Delta^{c} \bar{\Delta}^{c})+ \mu_{ij} 
  \textrm{Tr}(\tau_{2}\Phi^{T}_{i} \tau_{2} \Phi_{j}).
\end{eqnarray}
All couplings $Y^{(i)_{q,l}}$, $\mu_{ij}$, $\mu_{\Delta}$, $f$ in the
above potential, are complex with the the additional constraint that 
$\mu_{ij}$, $f$ and $f^{*}$ are symmetric matrices. It is clear from the 
above eq. that the theory has no baryon or lepton number violation 
terms. As such 
$R$-parity symmetry, defined by $(-1)^{3(B-L)+2S}$, is automatically 
conserved in the SUSYLR model.

It turns out that left-right symmetry imposes rather strong constraints on
the ground state of this model. It was pointed out by Kuchimanchi and
Mohapatra \cite{Kuchimanchi:1993jg} that there is no spontaneous parity
breaking for this minimal choice of Higgs in the supersymmetric left-right 
model and as such the ground state remains parity symmetric. If parity odd 
singlets are  introduced to break this symmetry \cite{Cvetic:1985zp}, then
it was shown \cite{Kuchimanchi:1993jg} that the charge-breaking vacua have a
lower potential than the charge-preserving vacua and as such the ground 
state does not conserve electric charge. Breaking $R$ parity was another 
possible solution to this dilemma of breaking parity symmetry. However, if 
one wants to prevent proton decay, then one must look for alternative 
solutions. One such possible solution is to add two new triplet superfields 
$\Omega(1,3,1,0)$, $\Omega_c (1,1,3,0)$ where under parity symmetry 
$\Omega \leftrightarrow \Omega_c^*$. This field has been explored 
extensively in \cite{Aulakh:1997ba,Aulakh:1997fq, Aulakh:1998nn, 
Yajnik:2006kc, Sarkar:2007er, Aulakh:1997vc}.

In the present paper we discuss another alternative solution with the
inclusion of a scalar bitriplet ($\eta$) and a parity odd singlet 
($\sigma$). This model breaks parity spontaneously, and also preserves 
electromagnetic charge automatically. The left-right parity is
spontaneously broken and as a result, the minimization does not
allow a left-right symmetry preserving solution.

\section{Supersymmetric Left-Right Symmetric Model including the
Bi-triplet and the Singlet}
\label{sec:bitrip-sing}

We now present our model, where we include a bi-triplet and
a parity odd singlet fields, in the minimal supersymmetric 
left-right symmetric model. These fields are vector-like and
hence do not contribute to anomaly, so we consider only one
of these fields. 
The quantum numbers for the new scalar fields $\eta$ and $\sigma$, under the 
gauge group considered are given by,
\begin{equation}
\eta(1,3,3,0), \qquad \sigma(1,1,1,0).
\end{equation}
Under parity, these fields transform as $\eta \leftrightarrow \eta$ and 
$\sigma \leftrightarrow -\sigma$. The superpotential for the model is
written in the more general tensorial notation,
\begin{eqnarray}
W &=& \nonumber f \eta_{\alpha i} \,\Delta_{\alpha} \,\Delta_{i}^{c} 
+ f^{*} \eta_{\alpha i} \,\bar{\Delta}_{\alpha} \,\bar{\Delta}_{i}^{c} 
\\ && \nonumber 
+~\lambda_1 \, \eta_{\alpha i} \, \Phi_{a m} \, \Phi_{b n} 
\left(\tau^{\alpha} \epsilon \right)_{a b} \left(\tau^{i} 
\epsilon \right)_{m n} + m_{\eta}\,\eta_{\alpha i}\, \eta_{\alpha i}
\\ && \nonumber 
+ ~M\, \left(\Delta_{\alpha} \bar{\Delta}_{\alpha}
+ \Delta_{i}^{c} \bar{\Delta}_{i}^{c} \right) 
+ \mu \, \epsilon_{a b}\, \Phi_{b m} \,\epsilon_{m n} \,\Phi_{a n} 
\\ && 
+~m_{\sigma}\, \sigma^{2} + \lambda_2 \, \sigma \left(\Delta_{\alpha} 
\bar{\Delta}_{\alpha} - \Delta_{i}^{c} \bar{\Delta}_{i}^{c} \right),
\end{eqnarray}
where, $\alpha$, $\beta$ $= 1, 2, 3$ and $a, b =1, 2$  are $SU(2)_{L}$ 
indices, whereas $i, j = 1, 2, 3$ and $m, n=1, 2$ are $SU(2)_{R}$ indices.
The summation over repeated index is implied, with the change in basis
from numerical $1,2,3$ indices to $+, -, 0$ indices as follows,
\begin{eqnarray}
\Psi_{\alpha} \Psi_{\alpha}  &=&\nonumber \Psi_{1} \Psi_{1}+\Psi_{2} \Psi_{2}
+\Psi_{3} \Psi_{3} \\ 
&=& \Psi_{+} \Psi_{-}+\Psi_{-} \Psi_{+}+\Psi_{0} \Psi_{0},
\end{eqnarray}
where, we have defined $\Psi_{\pm} =(\Psi_{1} \pm i \Psi_{2})/\sqrt{2}$ 
and $\Psi_{0}=\Psi_{3}$. The vacuum expectation values (vev) that the 
neutral components of the Higgs sector acquires are,
\begin{equation}
\begin{array}{lcl}
\langle \Delta_{-} \rangle = \langle \bar{\Delta}_{+} \rangle = v_{L},
&\qquad&
\langle \Delta_{+}^{c} \rangle = \langle \bar{\Delta}_{-}^{c} 
\rangle = v_{R}, \\[0.1cm]
\langle \Phi_{+ -} \rangle = v,  &\qquad&
\langle \Phi_{- +} \rangle = \it{v'},  \\
\langle \eta_{+ -} \rangle = \it{u_{1}}, \, &\qquad&
\langle \eta_{- +} \rangle = \it{u_{2}},\, \\
\langle \eta_{0 0} \rangle = \it{u_{0}}.
\end{array}
\end{equation}
Assuming SUSY to be unbroken till the TeV scale implies the $F$ and $D$ 
flatness conditions for the scalar fields to be,
\begin{eqnarray}
F_{\Delta_{\alpha}} &=& f \, \eta_{\alpha i} \,\Delta_{i}^{c} + M
\bar{\Delta}_{\alpha}+ \lambda_2 \, \sigma \,\bar{\Delta}_{\alpha} = 0,
\nonumber \\
F_{\bar{\Delta}_{\alpha}} &=& f^{*} \, \eta_{\alpha i} \,\bar{\Delta}_{i}^{c}
+M \Delta_{\alpha}  + \lambda_2 \, \sigma \,\Delta_{\alpha} = 0,
\nonumber \\
F_{\Delta_{i}^{c}} &=& f \, \eta_{\alpha i} \,\Delta_{\alpha} +M
\,\bar{\Delta}_{i}^{c} - \lambda_2 \, \sigma \,\bar{\Delta}^{c}_{i} = 0,
\nonumber \\ 
F_{\bar{\Delta}_{i}^{c}} &=& f^{*} \, \eta_{\alpha i} \,\bar{\Delta}_{i} +M
\,\Delta_{i}^{c} - \lambda_2 \, \sigma \,\Delta^{c}_{i}  = 0,
\nonumber \\
F_{\sigma} &=& 2 m_{\sigma}\, \sigma + \lambda_2 \,\left(\Delta_{\alpha}
\bar{\Delta}_{\alpha} - \Delta_{i}^{c} \bar{\Delta}_{i}^{c} \right) = 0,
\nonumber 
\end{eqnarray}
\begin{eqnarray}
F_{\eta_{\alpha i}}&=& f\, \Delta_{\alpha} \,\Delta_{i}^{c}
+f^{*} \bar{\Delta}_{\alpha} \,\bar{\Delta}_{i}^{c} 
+ 2\,m_{\eta}\,\eta_{\alpha i}
\nonumber \\ && 
+ ~\lambda_{1}  \, \Phi_{a m} \, \Phi_{b n} (\tau^{\alpha} \epsilon )_{a b}
(\tau^{i} \epsilon )_{m n}  = 0,
\nonumber 
\end{eqnarray}
\begin{eqnarray}
F_{\Phi_{c p}} &=& \lambda_{1} \,\eta_{\alpha i} \Phi_{b n} \,
(\tau^{\alpha} \epsilon )_{cb} \left(\tau^{i} \epsilon \right)_{pn} 
\nonumber \\ && 
+~ \lambda_{1} \,\eta_{\alpha i} \Phi_{a m} \,(\tau^{\alpha} \epsilon )_{a c}
(\tau^{i} \epsilon)_{mp}  
\nonumber  \\ && 
+ ~\mu \, \epsilon_{a c} \, \epsilon_{p n} \, \Phi_{a n} 
+ \mu \, \epsilon_{cb} \, \Phi_{b m} \,\epsilon_{mp} = 0,
\end{eqnarray}
\begin{eqnarray}
D_{R_{i}} &=& 2 {\Delta^{c}}^{\dagger} \tau_{i} \Delta^{c} + 2
\bar{\Delta}^{c \dagger} \tau_{i} \bar{\Delta^{c}}+  \eta \tau_{i}^{T}
\eta^{\dagger} +\Phi \tau_{i}^{T} \Phi^{ \dagger} =0, 
\nonumber \\
D_{L_{i}} &=& 2 \Delta^{ \dagger} \tau_{i} \Delta 
+ 2 \bar{\Delta}^{ \dagger} \tau_{i} \bar{\Delta}+\eta^{ \dagger} \tau_{i} 
\eta +\Phi^ { \dagger} \tau_{i} \Phi = 0,
\nonumber \\ 
D_{B-L} &=& 2 \left( \Delta^{ \dagger} \Delta -\bar{\Delta}^{ \dagger} 
\bar{\Delta}\right) - 2 \left( {\Delta^{c}}^{ \dagger} \Delta^{c} 
- \bar{\Delta}^{c \dagger}  \bar{\Delta}^{c}\right) = 0.
\end{eqnarray}
In the above eqns., we have neglected the slepton and squark fields, since
they would have zero vev at the scale considered.
We have also assumed $v' \ll v$ and hence the terms 
containing $v'$ can be neglected.

\section{Phenomenology} 
\label{sec:phenom}

An inspection of the minimisation conditions obtained at the end of the
previous section proves two important statements we have made earlier. First, the 
electromagnetic charge invariance of this vacuum is automatic for any 
parameter range of the theory. Secondly, the R-parity, defined in 
eq. (\ref{eq:rparity}), is preserved in the present model,
since the $\Delta$'s are R-parity even whereas the bi-doublet and the
bi-triplet Higgs scalars have zero R-parity. 

We shall now discuss the conditions that emerge from the vanishing of the 
various $F$ terms, which after the fields acquire their respective 
vevs, are given by,
\begin{eqnarray}
F_{\Delta} &=& f \, u_{1} v_{R} +(M+\lambda_{2} \langle \sigma 
    \rangle) v_L = 0, 
\label{eq:FDeltavev}
\\
F_{\bar{\Delta}} &=& 
f^{*} u_{2} v_{R}+(M+\lambda_{2} 
    \langle \sigma \rangle) v_{L} = 0, 
\label{eq:FDeltabarvev}
\\
F_{\Delta^{c}} &=& 
f\, u_{1} v_{L}+(M-\lambda_{2}\langle
    \sigma \rangle)v_{R} = 0, 
\label{eq:FDeltaCvev}
\\
F_{\bar{\Delta}^{c}} &=& 
f^{*}\, u_{2} v_{L}+(M-\lambda_{2}
    \langle \sigma \rangle) v_{R} = 0, 
\label{eq:FDeltaCbarvev}
\\
F_{\sigma} &=& m_{\sigma}\,\langle \sigma \rangle+ \lambda_2 (v_{L}^{2}
    -v_{R}^{2}) = 0, 
\label{eq:Fsigmavev}
\\
F_{\eta}  &=& 
f \, v_{L} v_{R} +f^{*} \, v_{L} 
    v_{R}+ \lambda_{1} v^{2} \,+ 2 m_{\eta}(u_{1}+u_{2}
    + u_{0}) = 0, 
\label{eq:FEtavev}
\\
F_{\Phi}  &=& - 2 \lambda_{1} (u_{1}+u_{2}) v
    + 2 \lambda_{1} u_{0} v-2 \mu v = 0.
\label{eq:FPhivev}
\end{eqnarray}
At the outset we see that the $F_\sigma$ flatness condition permits 
the trivial solution $\langle \sigma \rangle=0$, which would imply 
the undesirable solution $v_L=v_R$ and lead to no parity breakdown. 
But this special point can easily be destabilized once the soft
terms are turned on. Away from this special point, we are led 
to phenomenologically interesting vacuum  configurations.

The $F$ flatness conditions for the $\Delta$ and $\bar{\Delta}$
fields demand $fu_1=f^{*}u_2 $ which can be naturally satisfied by choosing  
\begin{equation}
f = f^{*} \qquad  \textrm{and} \qquad u_1 = u_2 \equiv u.
\end{equation}
This is consistent with
the relation obtained from the $F$ flatness conditions for the $\Delta^c$ and
$\bar{\Delta}^c$ fields, which may now be together read as
\begin{eqnarray}
(M-\lambda_{2} \langle \sigma \rangle)v_{R}=- f \, u v_{L}.
\label{eq:FDeltaBarCombo} 
\end{eqnarray}
The first four conditions (\ref{eq:FDeltavev})-(\ref{eq:FDeltaCbarvev}) 
can therefore be used to eliminate the scale $u$ and give a relation
\begin{equation}
\left( \frac{v_L}{v_R} \right)^2 = 
\frac{M-\lambda_{2} \langle \sigma \rangle}{M+\lambda_{2} 
\langle \sigma \rangle}.
\label{eq:ratio}
\end{equation}
Let us assume the scale of the vev's $u_1$, $u_2$ and $u_0$ to be 
the same. Then the
vanishing of $F_\eta$ gives a relation
\begin{equation}
 2f v_L v_R \approx - (\lambda_1 v^2 + 6 m_\eta u).
\label{eq:product}
\end{equation}
Finally, the last condition (\ref{eq:FPhivev}) has an interesting consequence. 
While electroweak symmetry is assumed to remain unbroken in the supersymmetric 
phase, so that $v$ must be chosen to be zero, we see that the factor 
multiplying  $v$ implies a relation 
\begin{equation}
 \mu \approx -\lambda_1 u.
\label{eq:musimu}
\end{equation}
 That is, taking $\lambda_1$ to be order unity, the scale of the 
$\mu$ term determines the scale of $u$. 

We now attempt an interpretation of these relations to obtain reasonable
phenomenology. The scale $v_R$ must be higher than the TeV scale. It seems
reasonable to assume that the eq. (\ref{eq:product}) provides a see-saw
relation between $v_L$ and $v_R$ vev's, and that this product is anchored
by the TeV scale. 
Since bitriplet contributes additional non-doublet Higgs in the 
Standard Model,
it is important that the vacuum expectation value $u$ is much higher
or much smaller than the electroweak scale, and we shall explore the
latter route. In this case $u$ should be strictly less than $1$GeV.  
The scale $m_\eta$ determines the masses of triplet majorons and 
needs to be high compared to the TeV scale. If the above see-saw relation
is not to be jeopardized, we must have $m_\eta u \leq m^2_{EW}$. We
can avoid proliferation of new mass scales by choosing
\begin{equation}
 m_\eta u \approx v^2 = m^2_{EW}.
\end{equation}
This establishes eq. (\ref{eq:product}) as the desired hierarchy 
equation, with $f$ chosen to be negative.

Now let us examine the consistency of the assumption $u \ll m_{EW}$
in the light of the two equations (\ref{eq:FDeltaBarCombo}) and 
(\ref{eq:ratio}). Let us assume that
$(v_L/v_R) \ll 1$ as in the non-supersymmetric case. 
Then eq. (\ref{eq:ratio}) means that on the right hand side,
\begin{equation}
M-\lambda_{2} \vev{\sigma} \ll M + \lambda_{2} \vev{\sigma}
\implies M \approx \lambda_{2} \langle \sigma \rangle.
\label{eq:Msigmaequality}
\end{equation}
Then eq. (\ref{eq:FDeltavev}) can be read as
\begin{equation}
\frac{v_L}{v_R} \approx \frac{(-f)u}{2M}.
\label{eq:ratiotwo} 
\end{equation}
We thus see that the required 
hierarchies of scales can be spontaneously generated, and can be related 
to each other. Finally, although only the ratios has been related in eq.
 (\ref{eq:ratiotwo}) we may choose
\begin{equation}
 v_L \approx u, \qquad v_R \approx M.
\label{eq:setscale}
\end{equation}
We see that through this choice of individual scales and through the see-saw
relation (\ref{eq:product}), $u$ and $v_R$ obey a mutual see-saw relation.
A small value of $u$ in the eV range would place $v_R$ in the intermediate
range as in the traditional proposals for neutrino mass see-saw. A larger
range of values close to the GeV scale would lead to $v_R$ and the
resulting heavy neutrinos states within the range of collider confirmation.

Finally, returning to eq. (\ref{eq:musimu}), we can obtain the desirable
scale for $u$ by choosing $\mu$ to be of that scale, viz., in
the sub-GeV range.
This solves the $\mu$ problem arising in MSSM by relating it to
other scales required to keep the $v_R$ high. 
An interesting consequence of the choices made so far is that using
eq.s (\ref{eq:Msigmaequality}) and (\ref{eq:setscale}) in eq. 
(\ref{eq:Fsigmavev}) yields
\begin{equation}
| m_\sigma | \approx \lambda_2\frac{v^2_{R}}{\langle \sigma \rangle} 
\sim \lambda_2^2 M.
\label{eq:msigmaM}
\end{equation}

To summarize, various phenomenological considerations lead to a
natural choice of three of the mass parameters of the superpotential,
$M$, $m_\sigma$ and $m_\eta$ to be comparable to each other and large, 
such as to determine $v_R$,
and in turn the masses of the heavy majorana neutrinos. The scale
$\mu$ which determines the vacuum expectation value $u$ and in turn
the value $v_L$ could be anything less than a GeV. Most importantly 
we have the see-saw relation eq. (\ref{eq:product}) which relates 
these scales, and if the $v_R$ scale is to be within a few orders 
of magnitude of the TeV scale, then $\mu$ should be close to though 
less than a GeV.

We can contemplate two extreme possibilities for the scale $M$. Keeping 
in mind the gravitino production and overabundance problem, we can choose
the largest value $v_R \leq 10^9$ GeV. If it can be ensured from
inflation that this is also the reheat temperature, then the 
thermalisation of heavy majorana neutrinos required for thermal
leptogenesis at a scale somewhat lower than this can be easily
accommodated. We can also try to take $v_R$ as low as
$10$ TeV which is consistent with preserving lepton asymmetry
generated by non-thermal mechanisms \cite{Sahu:2004sb}. 
Baryogenesis from non-thermal or sleptonic leptogenesis in 
this kind of setting has been extensively studied 
\cite{Grossman:2003jv, D'Ambrosio:2003wy, Boubekeur:2004ez, Chun:2005ms}. 
This low value of $v_R$ 
is consistent with neutrino see-saw relation, but will rely 
critically on the smallness of Yukawa  couplings\cite{Sahu:2004sb}
and may be accessible to colliders \cite{King:2004cx}. 

As we have seen, at the large scale, charge conservation also demands 
conservation of R-parity. The question generally arise as to what happens 
when heavy fields are integrated out and soft supersymmetry breaking terms 
are switched on. The analysis done in \cite{Aulakh:1997ba} implies 
that if $M_{R}$ 
is very large (around $10^{10}$ GeV), the breakdown of R-parity at low 
energy would give rise to an almost-massless majoron coupled to the Z-bosons, 
which is ruled out experimentally. This is one of the central aspects of 
supersymmetric left-right theories with large $M_{R}$: R-parity is an exact
symmetry of the low energy effective theory. The supersymmetric
partners of the neutrinos do not get any $vev$ at any scale,
which also ensures that the R-parity is conserved.

\section{Conclusion}
\label{sec:conclusion}

Supersymmetry and left-right symmetry are considered strong possible
candidates for extension of standard model. However, construction of a
low energy SUSYLR theory is by no means trivial, since left-right symmetry
cannot be broken spontaneously \cite{Kuchimanchi:1993jg}. In this paper, 
however, with the introduction of a bitriplet scalar field along with a 
parity odd Higgs singlet we have presented a possible mechanism of 
spontaneously breaking LR symmetry in a SUSYLR model. The advantages of
this model besides breaking parity spontaneously is that it preserves 
$R$ parity naturally. Also, we find a possible relation between the 
left-right symmetry breaking scale and the inverse of neutrino mass.

\section{Acknowledgments}
A.S. would like to thank the hospitality at PRL, where most of the present
work was done. US would like to
thank the Physics Department and the McDonnell Center for Space
Sciences, Washington University in St. Louis, USA for inviting him
as Clark Way Harrison visiting professor and thank R. Cowsik and F.
Ferrer for discussions.


\end{document}